\documentclass[preprint,12pt]{elsarticle}
\journal{International Journal of Plasticity}
\usepackage{graphics,epsfig}
\usepackage{graphicx}
\usepackage{float}
\usepackage{amssymb,amstext,amsmath}
\usepackage{mathtools,physics}
\usepackage{booktabs}
\usepackage{natbib}
\usepackage[latin1]{inputenc}
\usepackage{epstopdf}
\begin{document}
\newcommand{\balpha}{\boldsymbol{\alpha}}
\newcommand{\bbeta}{\boldsymbol{\beta}}
\newcommand{\bgamma}{\boldsymbol{\gamma}}
\newcommand{\bdelta}{\boldsymbol{\delta}}
\newcommand{\bepsilon}{\boldsymbol{\epsilon}}
\newcommand{\bvarepsilon}{\boldsymbol{\varepsilon}}
\newcommand{\bzeta}{\boldsymbol{\zeta}}
\newcommand{\bfoldeta}{\boldsymbol{\eta}}
\newcommand{\btheta}{\boldsymbol{\theta}}
\newcommand{\bvartheta}{\boldsymbol{\vartheta}}
\newcommand{\biota}{\boldsymbol{\iota}}
\newcommand{\bkappa}{\boldsymbol{\kappa}}
\newcommand{\blambda}{\boldsymbol{\lambda}}
\newcommand{\bmu}{\boldsymbol{\mu}}
\newcommand{\bnu}{\boldsymbol{\nu}}
\newcommand{\bxi}{\boldsymbol{\xi}}
\newcommand{\bpi}{\boldsymbol{\pi}}
\newcommand{\bvarpi}{\boldsymbol{\varpi}}
\newcommand{\brho}{\boldsymbol{\rho}}
\newcommand{\bvarrho}{\boldsymbol{\varrho}}
\newcommand{\bsigma}{\boldsymbol{\sigma}}
\newcommand{\bvarsigma}{\boldsymbol{\varsigma}}
\newcommand{\btau}{\boldsymbol{\tau}}
\newcommand{\bupsilon}{\boldsymbol{\upsilon}}
\newcommand{\bphi}{\boldsymbol{\phi}}
\newcommand{\bvarphi}{\boldsymbol{\varphi}}
\newcommand{\bchi}{\boldsymbol{\chi}}
\newcommand{\bpsi}{\boldsymbol{\psi}}
\newcommand{\bomega}{\boldsymbol{\omega}}
\newcommand{\bGamma}{\boldsymbol{\Gamma}}
\newcommand{\bDelta}{\boldsymbol{\Delta}}
\newcommand{\bTheta}{\boldsymbol{\Theta}}
\newcommand{\bLambda}{\boldsymbol{\Lambda}}
\newcommand{\bXi}{\boldsymbol{\Xi}}
\newcommand{\bPi}{\boldsymbol{\Pi}}
\newcommand{\bSigma}{\boldsymbol{\Sigma}}
\newcommand{\bUpsilon}{\boldsymbol{\Upsilon}}
\newcommand{\bPhi}{\boldsymbol{\Phi}}
\newcommand{\bPsi}{\boldsymbol{\Psi}}
\newcommand{\bOmega}{\boldsymbol{\Omega}}
\newcommand{\llbracket}{[\![}
\newcommand{\rrbracket}{]\!]}
\def\Xint#1{\mathchoice
   {\XXint\displaystyle\textstyle{#1}}%
   {\XXint\textstyle\scriptstyle{#1}}%
   {\XXint\scriptstyle\scriptscriptstyle{#1}}%
   {\XXint\scriptscriptstyle\scriptscriptstyle{#1}}%
   \!\int}
\def\XXint#1#2#3{{\setbox0=\hbox{$#1{#2#3}{\int}$}
     \vcenter{\hbox{$#2#3$}}\kern-.5\wd0}}
\def\ddashint{\Xint=}
\def\dashint{\Xint-}

\begin{frontmatter}
\title{Thermodynamic dislocation theory: Finite deformations} 
\author{K.C. Le\,$^{a,b}$\footnote{E-mail: lekhanhchau@tdtu.edu.vn}}
\address{$^a$\,Materials Mechanics Research Group, Ton Duc Thang University, Ho Chi Minh City, Vietnam\\
$^b$\,Faculty of Civil Engineering, Ton Duc Thang University, Ho Chi Minh City, Vietnam}

\begin{abstract}	
The present paper extends the thermodynamic dislocation theory initiated by Langer, Bouchbinder and Lookman [2010] to non-uniform finite plastic deformations. The equations of motion are derived from the variational equation involving the free energy density and the positive definite dissipation function. We also consider the simplified theory by neglecting the excess dislocations. For illustration, the problem of finite strain constrained shear of single crystals with one active slip system is solved within the proposed theory.
\end{abstract}

\end{frontmatter}

\section{Introduction}
\label{intro}

The novel approach initiated by \citet{Langer2010}, called LBL-theory for short (see also \citep{Langer2015,Langer2019}), has opened new perspectives in the dislocation mediated plasticity. Its main idea is to decouple the system of crystals containing dislocations into configurational and kinetic-vibrational subsystems. The configurational degrees of freedom describe the relatively slow, i.e. infrequent, atomic rearrangements associated with the irreversible motion of dislocations; the kinetic-vibrational degrees of freedom correspond to the fast oscillations of atoms about their equilibrium positions in the lattice. Due to the two different time scales characterizing these subsystems, two well-defined entropies (or temperatures) can be introduced. The governing equations of LBL-theory are based on the kinetics of thermally activated dislocation depinning and the irreversible thermodynamics of driven systems involving the configurational temperature. This LBL-theory has been successfully used to simulate the stress-strain curve for copper over fifteen decades in strain rates and for temperatures between room temperature and about one third of the melting temperature. Only a single fitting parameter is required to achieve full agreement with the experiments conducted by \citet{Follansbee1998} over a wide range of temperatures and strain rates. The theory was extended to include the interaction between two subsystems by \citet{Langer2016} and used to simulate the stress-strain curves for aluminum \citep{Le2017a,Le2017b} and steel \citep{Le2017b}, which capture the  thermal softening in full agreement with the experiments performed in \citep{Shi1997,Abbod2007}. The extension of the LBL-theory to non-uniform plastic deformations without including the excess dislocations was proposed by \citet{Langer2017}, \citet{LePiaoTr2018,LePiao2019} and was applied to the adiabatic shear band for steel and torsion of aluminum bars. The theory has again shown excellent agreement with the experiments conducted by \citet{Marchand1988}, \citet{Horstemeyer2002}, and \citet{Zhou1998}.

The LBL-theory and the above extensions can be used to describe the deformations of crystals whose dislocations are neutral in that their resultant Burgers vector vanishes. \citet{Ashby1970} called this kind of dislocations statistically stored, but we prefer the shorter and more precise name of redundant dislocations given earlier by \citet{Cottrell1964}. With non-uniform plastic deformations, like e.g. the torsion of a bar, the bending of a beam, or the deformation of polycrystals containing obstacles such as grain boundaries, solute atoms, or precipitates, redundant dislocations are accompanied by another type of dislocation to accommodate the plastic deformation gradient and ensure compatibility of the total deformation \citep{Nye1953,Bilby1955,Kroener1955,Ashby1970}. Most of contemporary experts in dislocation theory accept the proposal of \citet{Ashby1970} to call these dislocations ``geometrically necessary''. From the point of view of statistical mechanics of dislocations though (see e.g. \citep{Berdichevsky2006b,Limkumnerd2008,Poh2013,Zaiser2015}) the name excess dislocations seems more pregnant. Note that in recent years the density of excess dislocations can be measured indirectly by the high-resolution electron backscattering technique (EBSD) \citep{Kysar2010}. Although the percentage of excess dislocations in severly and plastically deformed crystals is low, they play a prominent role in the formation of the microstructure \citep{Ortiz1999,Huang2001,Kochmann2009a,Le2012,Koster2015,Koster2015a} and the Bauschinger and size effects \citep{Fleck1994,Nix1998,Hansen2004,Berdichevsky2007,Kochmann2009b,Kaluza2011,Le2013,Baitsch2015}. The continuum dislocation theory (CDT), which includes the density of excess dislocations as developed by \citet{Berdichevsky1967,Le1994,Le1996,Gurtin2002,Gurtin2005,Berdichevsky2006a,Le2014}, can actually capture microstructural formation and Bauschinger and size effects. Let us mention also the alternative approaches based on the discrete dislocation dynamics \citep{Giessen1995,Zbib2002}, the continuum dislocation dynamics  \citep{Hochrainer2014,Leung2015}, and the phase-field modeling \citep{Wang2001,Levitas2012}. However, the main drawback of these approaches to the dislocation mediated plasticity is the absence of redundant dislocations and configurational entropy (or its dual, configurational temperature). Since these two quantities are decisive for isotropic strain hardening \citep{Langer2010}, these approaches need a substantial revision. The thermodynamic dislocation theory, which accounts for redundant dislocations and configurational entropy, was first developed by \citet{Le2018} for small plastic deformations (see also \citep{LePiao2018a,LeTr2018}). It captures all experimentally observed features of the plastic flow: both the isotropic and kinematic hardening, the sensitivity of the stress-strain curves to temperature and strain rate, as well as the Bauschinger and size effects. Its application to the torsion of microwires \citep{LePiao2018} has shown an excellent agreement with the experiments of \citet{Liu2012} in the wide range of twist rates. 

The above TDT and its extensions have been proposed for small deformations. However, most of the experimental results were performed at finite plastic deformations. For logical consistency and comparison with experiments, the finite deformation TDT has to be developed. The aim of this paper is to provide the TDT for finite and non-uniform plastic deformations. By including its two missing quantities, configurational entropy (or its dual, configurational temperature) and density of redundant dislocations, as state variables in the constitutive equations, we obtain the synthesis of CDT and LBL-theory, which is a truly dynamic theory that applies to any non-uniform finite plastic deformations. We call it finite deformation thermodynamic dislocation theory. The latter is reduced to the previously proposed version of TDT if the total and plastic deformations are small. It is also consistent with the second law of thermodynamics. As an illustrative example, the problem of finite strain constrained shear of single crystal deforming in single slip is considered within the proposed theory. It will be shown that, for the specimens of macroscopic sizes, this problem can approximately be solved if the excess dislocations are neglected. The simulated stress-strain curves show temperature, velocity and orientation sensitivity.  

The paper is organized as follows. In Section 2 the finite deformation kinematics of TDT is laid down. Section 3 proposes the thermodynamic framework for this nonlinear TDT. In Section 4 the problem of finite strain constrained shear is analyzed. Section 5 presents the numerical solution of this problem and discusses the stress-strain curves and their sensitivity on temperature, strain rate, and orientation of slip system. Finally, Section 6 concludes the paper.  

\section{Finite deformation kinematics}
\label{sec:1}
For simplicity, we restrict ourselves to the plane strain deformations of a single crystal slab of thickness $L$ having only one active slip system. We choose a  fixed rectangular cartesian coordinate system associated with this slab such that the displacements $u_1$ and $u_2$ depend only on $x_1$ and $x_2$, but not on $x_3$, while the displacement $u_3=0$. The position vector of the same material point $(x_1,x_2,x_3)$ in the deformed state is
\begin{displaymath}
y_\alpha (x_\beta ,t)=x_\alpha +u_\alpha (x_\beta ,t), \quad y_3=x_3, \quad \alpha ,\beta =1,2.
\end{displaymath}
Thus, the deformation gradient is given by
\begin{equation}
\label{eq:1.0}
\vb{F}(x_\alpha ,t)=\vb{I}+\vb{u}\grad =\begin{pmatrix}
  1+u_{1,1}   &  u_{1,2}  &  0  \\
  u_{2,1}   &  1+u_{2,2}  &  0  \\
  0   &   0   &   1
\end{pmatrix},
\end{equation} 
where the comma in indices denotes partial differentiation. Kinematical quantities characterizing the observable deformation of this single crystal are the displacement field $\vb{u}(x_\alpha ,t)$ and the plastic deformation field $\vb{F}^p(x_\alpha ,t)$ that is in general incompatible. For single crystals having one active slip system, the plastic deformation is given by
\begin{equation}
\label{eq:1.1}
\vb{F}^p(x_\alpha ,t)=\vb{I}+\beta (x_\alpha ,t) \vb{s}\otimes \vb{m},
\end{equation}  
with $\beta (x_\alpha ,t)$ being the plastic slip, where the pair of constant and mutually orthogonal unit vectors $\vb{s}=(s_1,s_2,0)$ and $\vb{m}=(m_1,m_2,0)$ indicates the slip direction and the normal to the slip planes. The edge dislocations causing this plastic slip have dislocation lines parallel to the $x_3$-axis and the Burgers vector parallel to $\vb{s}$. Thus, there are altogether three degrees of freedom at each point of this generalized continuum. Later on, two additional internal variables will be introduced. In this paper, Greek indices run from 1 to 2, while Latin indices from 1 to 3. Summation over repeated indices within their range is understood. We see immediately from \eqref{eq:1.1} that $\det \vb{F}^p=1$, so the plastic deformation is volume preserving. 

\begin{figure}[htb]
	\centering
	\includegraphics[height=7cm]{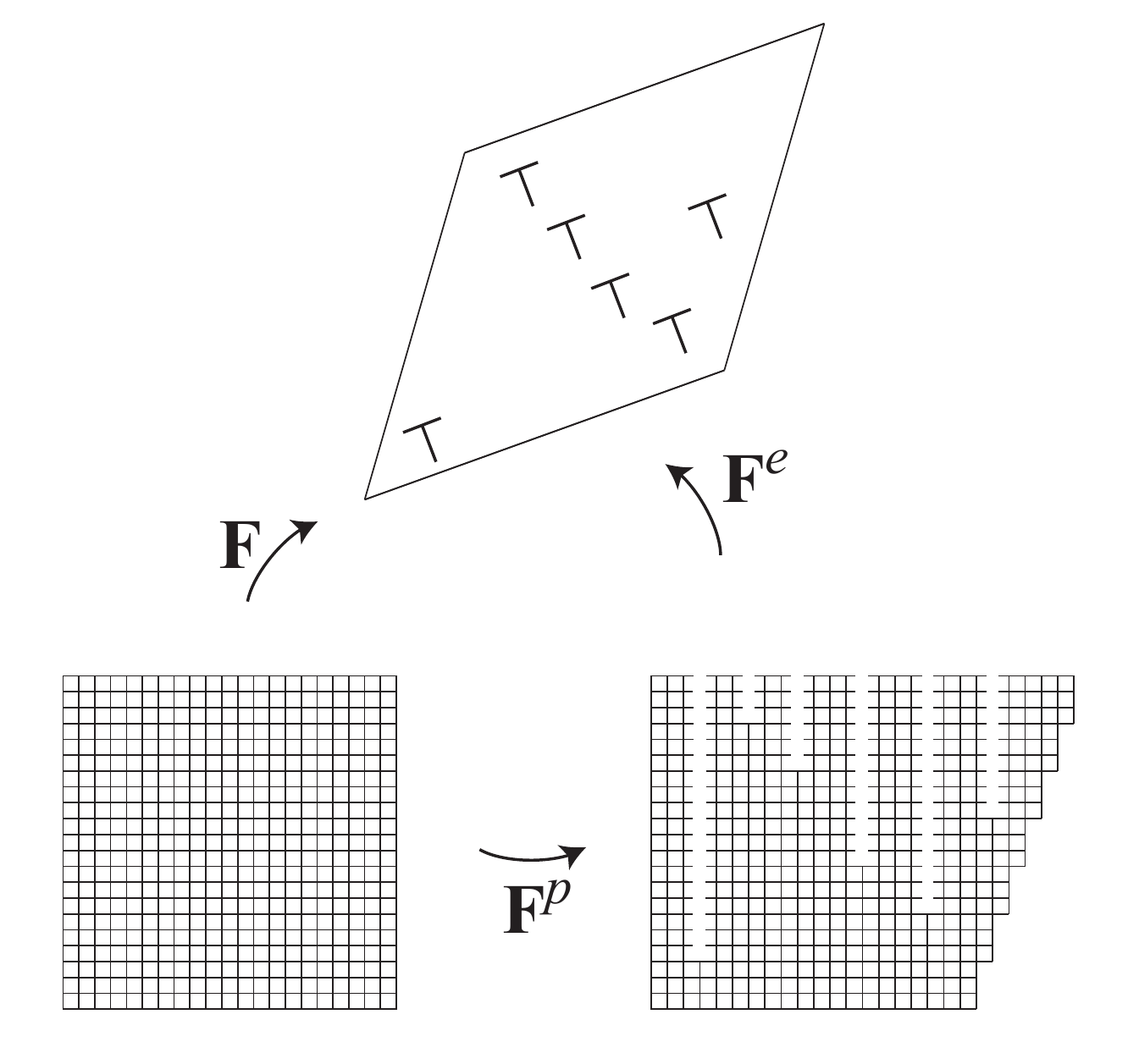}
	\caption{Multiplicative decomposition $\vb{F} =\vb{F}^e\cdot \vb{F}^p$}
	\label{fig:Nhslip}
\end{figure}

We assume that the deformation gradient $\vb{F}$ is decomposed into elastic and plastic part according to
\begin{equation*}
\vb{F} =\vb{F}^e\cdot \vb{F}^p.
\end{equation*}
Thus, the elastic deformation field equals
\begin{equation}
\label{eq:1.4}
\vb{F}^e=\vb{F}\cdot \vb{F}^{p-1}=\vb{F}\cdot (\vb{I}-\beta (x_\alpha ,t) \vb{s}\otimes \vb{m}).
\end{equation}
The relationship between the three deformation fields is illustrated schematically in Fig.~\ref{fig:Nhslip}. Looking at this Figure we see that the non-uniform plastic deformation $\vb{F}^p$ is that {\it creating} dislocations (either inside or at the boundary of the volume element) or {\it changing} their positions in the crystal without deforming the crystal lattice. In contrary, the elastic deformation $\vb{F}^e$ distorts the crystal lattice having {\it frozen} dislocations \citep{Le2014}.

With these deformation fields various measures of strain can be introduced. The relevant measures that will be used in TDT are the right Cauchy-Green deformation tensor
\begin{displaymath}
\vb{C}=\vb{F}^T\cdot \vb{F}
\end{displaymath}
and the similar elastic deformation tensor
\begin{displaymath}
\vb{C}^e=\vb{F}^{eT}\cdot \vb{F}^e.
\end{displaymath}
Note that both tensors are defined on the reference configuration \citep{Le2014}.

Consider a small representative area element $\dd a$ perpendicular to the $x_3$-axis through which a total number $N$ of dislocation lines cross. Assume that, among them, there are $N^+$ dislocations with the Burgers vector $b\, \vb{s}$ and $N^-$ dislocations with the Burgers vector $-b\, \vb{s}$, with $b$ being the magnitude of the Burgers vector. We define the number of excess dislocations as $N^g=|N^+-N^-|$ and the number of redundant dislocations as $N^r=N-N^g$. Correspondingly, the density of excess and redundant dislocations can be defined as
\begin{equation*}
\rho ^g=\frac{N^g}{\dd a}, \quad \rho ^r=\frac{N^r}{\dd a}.
\end{equation*}
It turns out that, under the plane strain deformation, the density of excess dislocations can be related to the plastic slip $\beta $ by the kinematic equation
\begin{equation}\label{excess}
\rho ^g=\frac{|\dd B|}{b\dd a}=\frac{1}{b}|\partial _s \beta |.
\end{equation}
This is based on the fact that the incompatibility tensor, introduced by \citet{Nye1953,Bilby1955,Kroener1955}, $\balpha=-\mathbf{F}^p \times \grad $, has only two non-zero components $\alpha _{i3}=s_i (\partial _s \beta )$, where $\partial _s=s_i\partial _i$ denotes the derivative in the direction $\mathbf{s}$. Thus, the resultant Burgers vector of all dislocations, whose lines cross the infinitesimal area $\mathrm{d}a$ perpendicular to the $x_3$-axis is
\begin{displaymath}
\mathrm{d}B_i=\alpha _{i3}\mathrm{d}a=s_i (\partial _s \beta )\mathrm{d}a.
\end{displaymath}
Since $|\mathrm{d}B|=b|N^+-N^-|$, the scalar density of excess dislocations is given by \eqref{excess}. This density can be indirectly measured by the high resolution electron backscatter diffraction technique (EBSD) \citep{Kysar2010}. 

In contrast to the excess dislocations, the other family of dislocations, called by \citet{Cottrell1964} and \citet{Weertman1996} redundant, cannot be expressed through the plastic distortion but nevertheless may have significant influences on the nucleation of excess dislocations and the work hardening of crystals. For any closed circuit surrounding an infinitesimal area (in the sense of continuum mechanics) the resultant Burgers vector of these dislocations always vanishes, so the closure failure caused by the incompatible plastic deformation is not affected by them. As a rule, the redundant dislocations in unloaded crystals at low temperatures exist in form of dislocation dipoles. The simple reason for this is that the energy of a dislocation dipole is much smaller than that of dislocations apart, so this bounded state of dislocations of opposite sign renders low energy to the whole crystal. From the other side, due to their low energy, the dislocation dipoles can be created by the mutual trapping of dislocations of different signs in a random way or, eventually, by thermal fluctuations in the presence of a stress field. Let us denote the density of redundant dislocations by $\rho ^r$. The total dislocation density is thus
\begin{equation*}
\rho =\rho ^g+\rho ^r.
\end{equation*}

\section{Thermodynamic dislocation theory}
\label{sec:2}

To set up phenomenological models of crystals with continuously distributed dislocations using the methods of non-equilibrium thermodynamics of driven system let us begin with the free energy density. As a function of the state, the free energy density may depend only on the state variables. Following \citet{Kroener1992} and \citet{Langer2016} we will assume that the elastic deformation tensor $\vb{C}^e=\vb{F}^{eT}\cdot \vb{F}^e$, the dislocation densities $\rho ^r$ and $\rho ^g$, the kinetic-vibrational temperature $T$, and the configurational temperature $\chi $ characterize the current state of the crystal, so these quantities are the state variables of the thermodynamic dislocation theory. The reason why the plastic deformation $\vb{F}^p$ cannot be qualified for the state variable is that it depends on the cut surfaces and consequently on the whole history of creating dislocations (for instance, climb or glide dislocations are created quite differently). Likewise, the gradient of plastic deformation tensor $\vb{C}^p=\vb{F}^{pT}\cdot \vb{F}^p$ cannot be used as the state variable by the same reason. In contrary, the dislocation densities $\rho ^r$ and $\rho ^g$ depend only on the characteristics of  dislocations in the current state (Burgers vector and positions of dislocation lines) and not on the history of their creation, so they are the proper state variable. We restrict ourself to the isothermal processes, so the kinetic-vibrational temperature $T$ is assumed to be constant and can be dropped in the list of arguments of the free energy density. Two state variables, $\vb{C}^e$ and $\rho ^g$ are dependent variables as they are expressible through the degrees of freedom $\vb{u}$ and $\beta$, while two others, $\rho ^r$ and $\chi $, can be regarded as independent internal variables. Our main assumption for the free energy density is
\begin{equation}\label{eq:2.1}
\psi (\vb{C}^e,\rho ^r,\rho ^g,\chi)= w(\vb{C}^e)+\gamma_D\rho ^r+\psi_m(\rho ^g) -\chi (- \rho \ln (a^2 \rho )+\rho )/L.
\end{equation}
The first term in \eqref{eq:2.1}, $w(\vb{C}^e)$, describes the stored energy density of crystal due to the elastic  deformation tensor $\vb{C}^e$. The second term is the self-energy density of redundant dislocations, with $\gamma_D$ being the energy of one dislocation per unit length. The third term is the energy density of excess dislocations. The last term has been introduced by \citet{Langer2016}, with $S_C=A(-\rho \ln (a^2 \rho )+\rho )$ having the meaning of the configurational entropy of dislocations, $A$ being the area occupied by the crystal slab.

We propose the following stored energy density of neo-Hookean compressible material
\begin{equation}
\label{eq:2.1a}
w(\vb{C}^e)=\frac{1}{2}\mu (I_1 -3)+\frac{1}{2}\lambda (1-J)^2-\mu \ln J,
\end{equation}
with $I_1=\tr \vb{C}^e$, $J=\det \vb{F}^e=\det \vb{F}$, and with $\lambda $ and $\mu $ being the Lame constants (cf. \citep{Kochmann2011}). It is easy to check that this formula reduces to the classical quadratic energy density of isotropic elastoplastic materials for small strain. Indeed, using Eqs.~\eqref{eq:1.0} and \eqref{eq:1.4} we find that, up to the cubic terms in $\vb{u}\grad $ and $\beta$,
\begin{equation*}
I_1=3+2\tr \bvarepsilon^e+\bvarepsilon^e \boldsymbol{:} \bvarepsilon^e +\bomega^e \boldsymbol{:} \bomega^e -2 \beta \, \vb{s}\cdot \vb{u}\grad \cdot \vb{m},
\end{equation*}
and
\begin{equation*}
J=1+\tr \bvarepsilon^e+\frac{1}{2}(\tr \bvarepsilon^e)^2-\frac{1}{2} \bvarepsilon^e \boldsymbol{:} \bvarepsilon^e +\frac{1}{2}\bomega^e \boldsymbol{:} \bomega^e - \beta \, \vb{s}\cdot \vb{u}\grad \cdot \vb{m},
\end{equation*}
where $\bvarepsilon^e=\frac{1}{2}(\vb{u}\grad +\grad \vb{u}-\beta \vb{s}\otimes \vb{m}-\beta \vb{m}\otimes \vb{s})$ is the small elastic strain tensor and $\bomega^e=\frac{1}{2}(\vb{u}\grad -\grad \vb{u}-\beta \vb{s}\otimes \vb{m}+\beta \vb{m}\otimes \vb{s})$ the small elastic rotation tensor. Expanding function $f(J)=\frac{1}{2}\lambda (1-J)^2-\mu \ln J$ into the Taylor series about $J=1$ and keeping in Eq.~\eqref{eq:2.1a} only the quadratic terms, we reduce it to the classical formula
\begin{displaymath}
w(\bvarepsilon^e)=\frac{1}{2}\lambda (\tr \bvarepsilon^e)^2+\mu \bvarepsilon^e \boldsymbol{:} \bvarepsilon^e.
\end{displaymath}
Furthermore, we employ the following formula for the energy of excess dislocations $\psi_m(\rho ^g)$
\begin{equation*}
\psi_m(\rho ^g)=\frac{\gamma_D}{a^2}\ln \frac{1}{1-a^2\rho ^g}.
\end{equation*}
This logarithmic energy term stems from two facts: (i) energy of excess dislocations for small dislocation densities must be $\gamma_D\rho ^g$ like that of redundant dislocations, and (ii) there exists a saturate state of maximum disorder and infinite configurational temperature characterized by the admissibly closest distance between excess dislocations, $a$. The logarithmic term \citep{Berdichevsky2006b} ensures a linear increase of the energy for small dislocation density $\rho ^g$ and tends to infinity as $\rho ^g$ approaches the saturated dislocation density $1/a^2$ hence providing an energetic barrier against over-saturation.

With this free energy density we can now write down the energy functional of the dislocated crystal. Let the cross section of the undeformed single crystal occupy the region $\mathcal{A}$ of the $(x_1,x_2)$-plane. The boundary of this region, $\partial \mathcal{A}$, is assumed to be the closure of union of two non-intersecting curves, $\partial _k$ and $\partial _s$. Let the displacement vector $\vb{u}_(\mathbf{x},t)$ be a given smooth function of coordinates (clamped boundary), and, consequently, the plastic slip $\beta (\vb{x},t)$ vanishes 
\begin{equation}\label{eq:2.2}
\vb{u}(\vb{x},t)=\tilde{\vb{u}}(\vb{x},t),\quad \beta (\vb{x},t)=0 \quad \text{for $\vb{x}\in \partial _k$}.
\end{equation}
The remaining part $\partial _s$ of the boundary is assumed to be traction-free. If no body force acts on this crystal, then its energy functional per unit depth is defined as
\begin{equation*}
I[\vb{u}(\vb{x},t),\beta (\vb{x},t),\rho ^r(\vb{x},t),\chi (\vb{x},t))]=\int_{\mathcal{A}}\psi (\vb{C}^e,\rho ^r, \rho ^g,\chi )\dd{a},
\end{equation*}
with $\dd a=\dd x_1\dd x_2$ denoting the area element. 

Under the increasing load the resolved shear stress also increases, and when it reaches the Taylor stress, dislocations dipoles dissolve and begin to move until they are trapped again by dislocations of opposite sign. During this motion dislocations always experience the resistance causing the energy dissipation. The increase of dislocation density as well as the increase of configurational temperature also lead to the energy dissipation. Neglecting the dissipation due to internal viscosity associated with the strain rate, we propose the dissipation potential in the form 
\begin{equation}\label{eq:2.4}
D(\dot{\beta},\dot{\rho},\dot{\chi})=\tau _Y \dot{\beta }+\frac{1}{2}d_\rho \dot{\rho }^2+\frac{1}{2}d_\chi \dot{\chi }^2,
\end{equation}
where $\tau_Y$ is the flow stress during the plastic yielding, $d_\rho $ and $d_\chi $ are still unknown functions, to be determined later. The first term in \eqref{eq:2.4} is the plastic power which is assumed to be homogeneous function of first order with respect to the plastic slip rate \citep{Puglisi2005}. The other two terms describe the dissipation caused by the multiplication of dislocations and the increase of configurational temperature \citep{Langer2010}. 

Since the dislocation mediated plastic flow is the irreversible process, we derive the governing equations from the following variational principle: the true displacement field $\check{\mathbf{u}}(\mathbf{x},t)$, the true plastic slips $\check{\beta }(\mathbf{x},t)$, the true density of redundant dislocations $\check{\rho }^r(\mathbf{x},t)$, and the true configurational temperature $\check{\chi }(\mathbf{x},t)$ obey the variational equation
\begin{equation}
\label{eq:2.8}
\delta I+\int_{\mathcal{A}} \left( \frac{\partial D}{\partial \dot{\beta }}\delta \beta +\frac{\partial D}{\partial \dot{\rho }}\delta \rho +\frac{\partial D}{\partial \dot{\chi }}\delta \chi \right) \dd{a}=0
\end{equation}
for all variations of admissible fields $\vb{u}(\vb{x},t)$, $\beta (\vb{x},t)$, $\rho ^r(\vb{x},t)$, and $\chi (\vb{x},t)$ satisfying the constraints \eqref{eq:2.2}. 

Varying the energy functional with respect to $\vb{u}$ we obtain the quasi-static equations of equilibrium of macro-forces
\begin{equation}
\label{eq:2.9}
\vb{T}\cdot \grad =0, \quad \vb{T}=\frac{\partial \psi}{\partial \vb{F}}=\mu \vb{F}^e\cdot \vb{F}^{p-T}-[\lambda (1-J)J+\mu]\vb{F}^{-T}, 
\end{equation}
which are subjected to the boundary conditions \eqref{eq:2.2} and 
\begin{equation*}
\vb{T}\cdot \vb{n}=0 \quad \text{on $\partial _s$}.
\end{equation*}
Taking the variation of $I$ with respect to three other quantities $\beta $, $\rho ^r$, and $\chi $ and requiring that Eq.~\eqref{eq:2.8} is satisfied for their admissible variations, we get three equations
\begin{equation}
\label{eq:2.10}
\begin{split}
\tau -\tau_B -\tau _Y=0, 
\\
(e_D+\chi \ln (a^2\rho))/L +d_\rho \dot{\rho }=0, 
\\
(\rho \ln (a^2\rho) -\rho)/L+d_\chi \dot{\chi }=0,
\end{split}
\end{equation}
with $e_D=\gamma_DL$. Here,
\begin{equation*}
\tau =-\frac{\partial \psi}{\partial \beta }=\mu \, \vb{s}\cdot (\vb{C}\cdot \vb{F}^{p-1})\cdot \vb{m}
\end{equation*}
is the resolved shear stress (Schmid stress), while 
\begin{equation*}
\tau_{B}=-\frac{\partial^2\psi_m}{\partial (\rho^g)^2}\beta_{,ss}
\end{equation*}
is the back stress. The first equation of \eqref{eq:2.10} can be interpreted as the balance of microforces acting on dislocations. This equation is subjected to the Dirichlet boundary condition \eqref{eq:2.2}$_2$ on $\partial _k$ and
\begin{displaymath}
\frac{\partial \psi_m}{\partial \rho^g}=\gamma_D \quad \text{on $\partial _s$}.
\end{displaymath}

\citet{Le2018,LePiao2018} have shown that Eqs.~\eqref{eq:2.10}$_{2,3}$ reduces to the corresponding equations for $\chi $ and $\rho $ of LBL-theory describing the motion of the system driven by the constant shear rate $\dot{\gamma}=q_0/t_0$ \citep{Langer2010}
\begin{equation*}
\begin{split}
\dot{\chi } =  K_\chi \frac{\tau q(\tau ,\rho )}{\mu t_0} \left[ 1-\frac{\chi }{\chi ^{ss}(q)} \right], 
\\
\dot{\rho } = K_\rho \frac{\tau }{a^2\mu \nu (\theta,\rho ,q_0)^2}\frac{q(\tau ,\rho )}{t_0}\left[ 1-\frac{\rho }{\rho ^{ss}(\chi )} \right] , 
\end{split}
\end{equation*}
if we choose
\begin{equation}
\begin{split}
\label{eq:2.12}
d_\chi =\frac{\rho -\rho \ln (a^2\rho) }{LK_\chi \frac{\tau_Y q(\tau_Y ,\rho )}{\mu t_0} \left[ 1-\frac{\chi }{\chi _0} \right] },
\\
d_\rho =\frac{-e_D-\chi \ln (a^2\rho)}{LK_\rho \frac{\tau_Y }{a^2\mu \nu (\theta,\rho ,q_0)^2}\frac{q(\tau_Y ,\rho )}{t_0}\left[ 1-\frac{\rho }{\rho ^{ss}(\chi )} \right]} ,
\end{split}
\end{equation}
where $t_0$ is the characteristic microscopic time scale. In these equations the steady-state configurational temperature is denote by $\chi _0$, while the steady-state dislocation density equals
\begin{displaymath}
\rho ^{ss}(\chi )=\frac{1}{a^2}e^{-e_D/\chi }.
\end{displaymath}
Finally, $\nu (\theta,\rho ,q_0)$ is defined as follows
\begin{equation*}
\nu (\theta,\rho ,q_0)=\ln \left( \theta \right) - \ln \left[ \frac{1}{2}\ln \left( \frac{b^2\rho }{q_0^2}\right) \right] .
\end{equation*}
Note that, for $\rho $ changing between 0 and $\rho ^{ss}<1/a^2$, both numerators on the right-hand sides of \eqref{eq:2.12} are positive, and the dissipative potential \eqref{eq:2.4} is positive definite as required by the second law of thermodynamics.

To close this system an evolution equation for $\tau_Y$ is required. We consider first the uniform deformation and take the time derivative of the equation for the resolved shear stress 
\begin{equation}
\label{eq:2.5a}
\dot{\tau }=\mu \, \vb{s}\cdot (\dot{\vb{C}}\cdot \vb{F}^{p-1}+\vb{C}\cdot \dot{\vb{F}}^{p-1})\cdot \vb{m}=\mu [\vb{s}\cdot (\dot{\vb{C}}\cdot \vb{F}^{p-1}) \cdot \vb{m}-\dot{\beta }\, \vb{s}\cdot \vb{C}\cdot \vb{s}].
\end{equation}
Based on the Orowan's equation $\dot{\beta}=b \rho v$ and the assumption that dislocations spend most of time in the pinned state, the plastic slip rate must be determined by the kinetics of thermally activated dislocation depinning \citep{Langer2010}. This yields 
\begin{equation}
\label{eq:2.5}
\dot{\beta}=\frac{q(\tau,\rho)}{t_0},\quad q(\tau,\rho)=b\sqrt{\rho } [f_P(\tau,\rho)-f_P(-\tau,\rho)],
\end{equation}
where
\begin{equation}
\label{exp}
f_P(\tau,\rho)=\exp\,\Bigl[-\,\frac{1}{\theta}\,e^{-\tau/\tau_T(\rho)}\Bigr]. 
\end{equation}
In Eq.~\eqref{exp} the dimensionless temperature is introduced as $\theta =T/T_P$, with $T_P$ being the pinning energy barrier, while $\tau _T=\mu _T b\sqrt{\rho }$ is the Taylor stress. Substituting \eqref{eq:2.5} into \eqref{eq:2.5a}, we obtain the evolution of the resolved shear stress for the uniform plastic deformation. However, for non-uniform plastic deformations producing excess dislocations $q(\gamma)/t_0$ does not equal $\dot{\beta}$, and we associate $q(\gamma)/t_0$ to the plastic shear rate caused by the depinning of redundant dislocations only. In this case the equation for the flow stress in rate form is proposed as follows
\begin{equation}
\label{eq:2.6}
\dot{\tau }_Y=\mu \Bigl[ \vb{s}\cdot (\dot{\vb{C}}\cdot \vb{F}^{p-1}) \cdot \vb{m}-\frac{q(\tau_Y,\rho)}{t_0}\, \vb{s}\cdot \vb{C}\cdot \vb{s}\Bigr] .
\end{equation}
Eqs.~\eqref{eq:2.9} and \eqref{eq:2.10}, combined with \eqref{eq:2.5} and \eqref{eq:2.6}, yield the  equations of motion of dislocated crystal. Note that, for the fast loading when the inertia term becomes essential, Eq.~\eqref{eq:2.9} should be modified to
\begin{equation*}
\varrho_0 \ddot{\vb{u}}=\vb{T}\cdot \grad ,  
\end{equation*}
with $\varrho_0 $ being the initial mass density.

For uniform deformations the theory is considerably simplified. Indeed, since $\rho^g=0$, the back stress $\tau_B$ vanishes, and Eq.~\eqref{eq:2.10}$_1$ implies that $\tau =\tau_Y$. Besides, as the equilibrium of macro-forces \eqref{eq:2.9} is satisfied automatically, the whole system reduces to
\begin{equation}
\begin{split}
\dot{\beta}=\frac{q(\tau,\rho)}{t_0}, 
\\
\dot{\tau }=\mu \Bigl[ \vb{s}\cdot (\dot{\vb{C}}\cdot \vb{F}^{p-1}) \cdot \vb{m}-\frac{q(\tau,\rho)}{t_0}\, \vb{s}\cdot \vb{C}\cdot \vb{s}\Bigr] ,
\\
\dot{\chi } =  K_\chi \frac{\tau q(\tau ,\rho )}{\mu t_0} \left[ 1-\frac{\chi }{\chi _0} \right], \label{eq:2.13}
\\
\dot{\rho } = K_\rho \frac{\tau }{a^2\mu \nu (\theta ,\rho ,q_0)^2}\frac{q(\tau ,\rho )}{t_0}\left[ 1-\frac{\rho }{\rho ^{ss}(\chi )} \right] .
\end{split}
\end{equation}

\section{Finite strain constrained shear}
\label{sec:3}

\begin{figure}[htb]
	\centering
	\includegraphics[height=7cm]{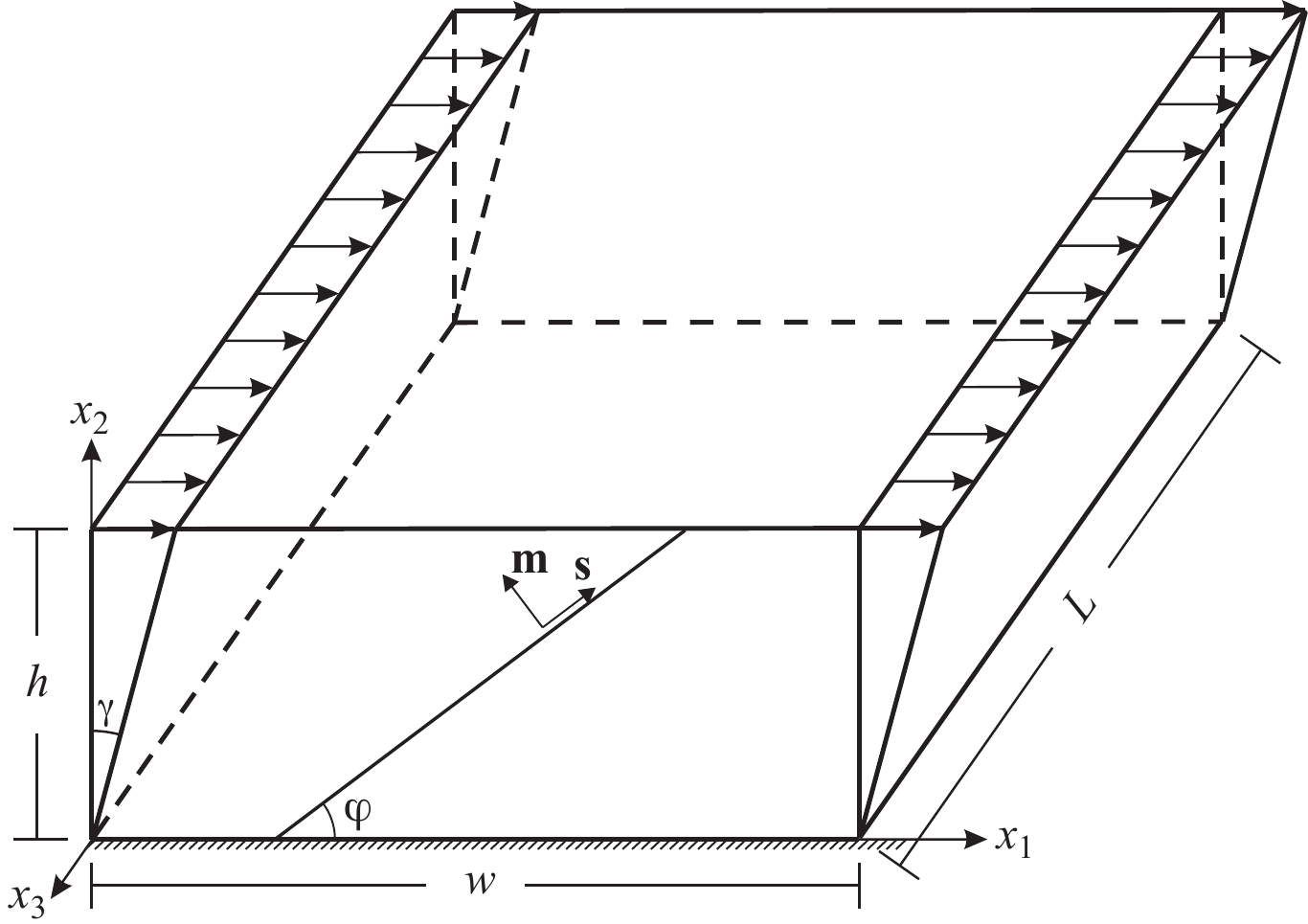}
	\caption{Finite strain constrained shear}
	\label{fig:planeshear}
\end{figure}

As an application of the proposed theory let us consider the single crystal layer having a rectangular cross-section of width $w$ and height $h$, $0 \leq x \leq w$, $0 \leq y \leq h$ and undergoing a finite plane strain constrained shear deformation (see Fig.~\ref{fig:planeshear}). The single crystal is placed in a hard device with prescribed displacements at its upper and lower sides as
\begin{equation}\label{SS_boundary}
    u_1(0,t)=0, \quad u_2(0,t)=0, \quad u_1(h,t)=\gamma(t) h, \quad u_2(h,t)=0,
\end{equation}
where $u_1(y,t)$ and $u_2(y,t)$ are the longitudinal and transverse displacements, respectively, with $\gamma(t)$ being the overall shear regarded as a control parameter. We assume the shear rate $\dot{\gamma}=q_0/t_0$ to be constant. If $\gamma $ is small, the layer deforms elastically. However, when $\gamma $ becomes sufficiently large, dislocations will be depinned initiating the plastic flow. We admit only one active slip system, with the slip directions inclined at an angle $\varphi $ to the $x_1$-axis and the edge dislocations whose lines are parallel to the $x_3$-axis. Under these conditions the deformation can be assumed uniform for the specimens of macroscopic sizes. Strictly speaking, the conditions \eqref{SS_boundary} do not allow dislocations to reach the upper and lower boundaries, so these boundaries act as obstacles. This leads to an accumulation of excess dislocations and to nonuniformity of plastic deformation near these boundaries. For macroscopic specimens, however, the percentage of excess dislocations is negligibly low and the assumption of uniformity of plastic deformation is a good approximation everywhere except for the very thin layers near the upper and lower boundaries.  We aim at determining the displacements, the dislocation densities, the configuration temperature and the stress-strain curve as a function of $\gamma $ within the finite deformation thermodynamic dislocation theory summarized in \eqref{eq:2.13}. 

Under these conditions the deformation gradient and the right Cauchy-Green deformation tensor are given by
\begin{equation} \label{total}
\vb{F}=\begin{pmatrix}
  1   &  \gamma  &  0  \\
  0   &  1  &  0  \\
  0   &   0   &   1
\end{pmatrix}, \quad \vb{C}=\vb{F}^T\cdot \vb{F}=\begin{pmatrix}
  1   &  \gamma  &  0  \\
  \gamma   &  1+\gamma^2  &  0  \\
  0   &   0   &   1
\end{pmatrix}.
\end{equation}
The active slip system inclined at the angle $\varphi $ to the $x_1$-axis has the vectors  $\mathbf{s}=(\cos \varphi , \sin \varphi ,0)$ and $\mathbf{m}=(-\sin \varphi , \cos \varphi ,0)$, therefore the plastic deformation and its inverse are
\begin{equation}
\begin{split}
\vb{F}^p=
    \begin{pmatrix}
        1-\beta \sin \varphi \cos \varphi & \beta \cos ^{2} \varphi  & 0 \\
        -\beta \sin ^{2} \varphi  & 1+\beta \sin \varphi \cos \varphi  & 0 \\
        0 & 0 & 1
        \end{pmatrix},
        \\ 
\vb{F}^{p-1}=
    \begin{pmatrix}
        1+\beta \sin \varphi \cos \varphi & -\beta \cos ^{2} \varphi  & 0 \\
        \beta \sin ^{2} \varphi  & 1-\beta \sin \varphi \cos \varphi  & 0 \\
        0 & 0 & 1
        \end{pmatrix}. \label{plastic}
\end{split}
\end{equation}

Because the system is undergoing steady-state plane constrained shear with the constant shear rate $\dot{\gamma}=q_0/t_0$, we can replace the time $t$ by the total amount of shear $\gamma$ so that $t_0\dv*{}{t} \to q_0\dv*{}{\gamma}$. The equations of motion for this system become
\begin{equation}
\begin{split}
\dv{\beta}{\gamma}=\frac{q(\tau,\rho)}{q_0}, 
\\
\dv{\tau }{\gamma}=\mu \Bigl[ \vb{s}\cdot (\dv{\vb{C}}{\gamma}\cdot \vb{F}^{p-1}) \cdot \vb{m}-\frac{q(\tau,\rho)}{q_0}\, \vb{s}\cdot \vb{C}\cdot \vb{s}\Bigr] ,
\\
\dv{\chi }{\gamma} =  K_\chi \frac{\tau q(\tau ,\rho )}{\mu q_0} \left[ 1-\frac{\chi }{\chi _0} \right], \label{eq:3.1}
\\
\dv{\rho }{\gamma} = K_\rho \frac{\tau }{a^2\mu \nu (\theta,\rho ,q_0)^2}\frac{q(\tau ,\rho )}{q_0}\left[ 1-\frac{\rho }{\rho ^{ss}(\chi )} \right] .
\end{split}
\end{equation}
With $\vb{C}$ from \eqref{total} and $\vb{F}^{p-1}$ from \eqref{plastic} we find that
\begin{equation*}
\begin{split}
f_1(\beta ,\gamma ,\varphi )\equiv \vb{s}\cdot (\dv{\vb{C}}{\gamma}\cdot \vb{F}^{p-1}) \cdot \vb{m}=-\beta \gamma +(1+\beta \gamma )\cos 2\varphi +(\gamma -\beta )\sin 2\varphi ,
\\
f_2(\gamma ,\varphi )\equiv \vb{s}\cdot \vb{C}\cdot \vb{s} = 1+\gamma ^2\sin^2 \varphi + \gamma \sin 2\varphi .
\end{split}
\end{equation*}
In terms of the introduced function Eq.~\eqref{eq:3.1}$_{2}$ can be written as follows
\begin{equation*}
\dv{\tau }{\gamma}=\mu \Bigl[ f_1(\beta ,\gamma ,\varphi )-f_2(\gamma ,\varphi )\frac{q(\tau,\rho)}{q_0}\Bigr] .
\end{equation*}

It is convenient to rewrite Eqs.~\eqref{eq:3.1} in terms of the following dimensionless variables 
\begin{equation*}
\tilde{\rho}=a^2\rho , \quad 
\tilde{\chi }=\frac{\chi }{e_D}.
\end{equation*}
Then we present Eq.~(\ref{eq:2.5}) in the form
\begin{equation*}
q(\tau,\rho)=\frac{b}{a}\tilde{q}(\tau,\tilde{\rho}),
\end{equation*}
where
\begin{equation*}
\tilde{q}(\tau,\tilde{\rho})=\sqrt{\tilde{\rho}}[\tilde{f}_P(\tau,\tilde{\rho})-\tilde{f}_P(-\tau,\tilde{\rho})].
\end{equation*}
We set $\tilde{\mu}_T=(b/a)\mu_T=\mu r$ and assume that $r$ is independent of temperature and strain rate. Then
\begin{equation*}
\tilde{f}_P(\tau,\tilde{\rho})=\exp\,\Bigl[-\,\frac{1}{\theta}\,e^{-\tau/(\mu r\sqrt{\tilde{\rho }})}\Bigr].
\end{equation*}
We define $\tilde{q}_0=(a/b)q_0$ so that $q/(q_0)=\tilde{q}/\tilde{q}_0$. Function $\nu $ becomes
\begin{equation*}
\tilde{\nu}(\theta,\tilde{\rho},\tilde{q}_0) \equiv \ln\Bigl(\frac{1}{\theta}\Bigr) - \ln\Bigl[\ln\Bigl(\frac{\sqrt{\tilde{\rho}}}{\tilde{q}_0}\Bigr)\Bigr].
\end{equation*}
The dimensionless steady-state quantities are
\begin{equation*}
\tilde{\rho}^{ss}(\tilde{\chi})=e^{-1/\tilde{\chi}}, \quad \tilde{\chi}_0=\chi_0/e_D.
\end{equation*}
Using $\tilde{q}$ instead of $q$ as the normalized plastic strain rate means that we are effectively rescaling $t_0$ by a factor $b/a$. Since $t_0^{-1}$ is a microscopic attempt frequency, of the order $10^{12}$\,s$^{-1}$, we take $(a/b)t_0=10^{-12}$s.

In terms of the introduced quantities the governing equations read
\begin{equation}\label{eq:3.5}
\begin{split}
\dv{\beta}{\gamma}=\frac{q(\tau,\rho)}{q_0}, 
\\
\dv{\tau}{\gamma} = \mu \Bigl[ f_1(\beta ,\gamma ,\varphi )-f_2(\gamma ,\varphi )\frac{\tilde{q}(\tau,\tilde{\rho})}{\tilde{q}_0}\Bigr] ,  
\\
\dv{\tilde{\chi }}{\gamma} =  K_\chi \frac{\tau \tilde{q}(\tau ,\tilde{\rho })}{\mu \tilde{q}_0} \left[ 1-\frac{\tilde{\chi }}{\tilde{\chi }_0} \right], 
\\
\dv{\tilde{\rho }}{\gamma} = K_\rho \frac{\tau }{\mu \tilde{\nu }(\theta ,\tilde{\rho },\tilde{q}_0)^2}\frac{\tilde{q}(\tau ,\tilde{\rho })}{\tilde{q}_0}\left[ 1-\frac{\tilde{\rho }}{\tilde{\rho }^{ss}(\chi )} \right] .
\end{split}
\end{equation}
The corresponding equations of the small strain theory are
\begin{equation}\label{eq:3.6}
\begin{split}
\dv{\beta}{\gamma}=\frac{q(\tau,\rho)}{q_0}, 
\\
\dv{\tau}{\gamma} = \mu \Bigl[ \cos 2\varphi -\frac{\tilde{q}(\tau,\tilde{\rho})}{\tilde{q}_0}\Bigr] ,  
\\
\dv{\tilde{\chi }}{\gamma} =  K_\chi \frac{\tau \tilde{q}(\tau ,\tilde{\rho })}{\mu \tilde{q}_0} \left[ 1-\frac{\tilde{\chi }}{\tilde{\chi }_0} \right], 
\\
\dv{\tilde{\rho }}{\gamma} = K_\rho \frac{\tau }{\mu \tilde{\nu }(\theta ,\tilde{\rho },\tilde{q}_0)^2}\frac{\tilde{q}(\tau ,\tilde{\rho })}{\tilde{q}_0}\left[ 1-\frac{\tilde{\rho }}{\tilde{\rho }^{ss}(\chi )} \right] .
\end{split}
\end{equation}

\section{Numerical simulations}
\label{sec:4}

\begin{figure}[htb]
	\centering
	\includegraphics[height=5cm]{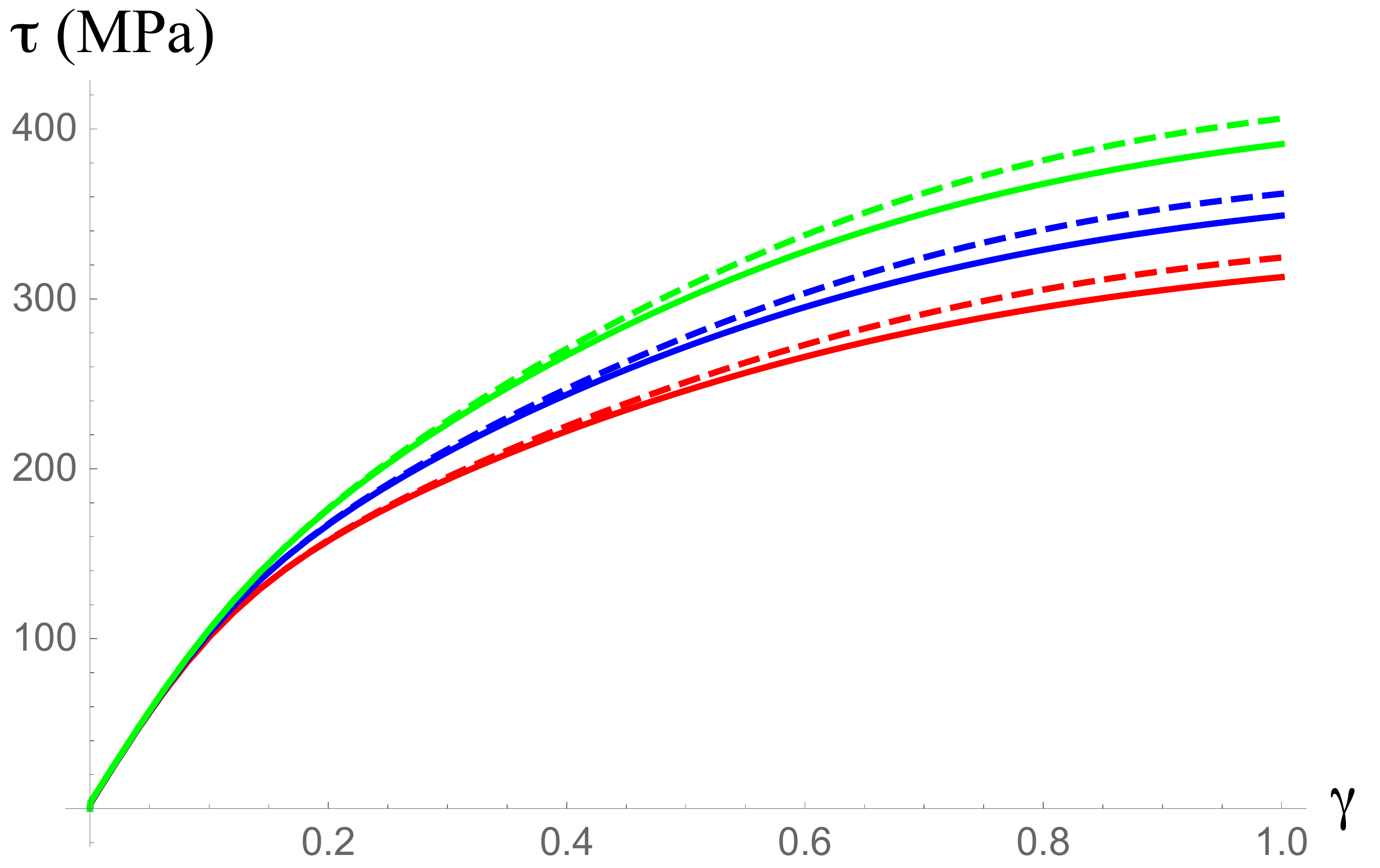}
	\caption{Resolved shear stress $\tau $ (in MPa) versus shear strain $\gamma $ for copper at the room temperature $298\,$K, at $\varphi =30^\circ$, and at three shear rates $0.1/$s (red), $10/$s (blue), and $1000/$s (green): (i) finite strain (bold line), (ii) small strain theory (dashed line).}
	\label{fig:3}
\end{figure}

In order to simulate the theoretical stress-strain curves, we need values for five system-specific parameters and two initial conditions from each sample.  The five basic parameters are the following: the activation temperature $T_P$, the stress ratio $r$, the steady-state scaled effective temperature $\tilde\chi_0$, and the two dimensionless conversion factors $K_\chi$ and $K_\rho$. We also need initial values of the scaled dislocation density $\tilde\rho_i$ and the effective disorder temperature $\tilde\chi_i$; all of which are determined by the sample preparation. The basic parameters for copper are chosen as follows \citep{Langer2010} 
\begin{equation*}
T_P=40800\, \text{K}, r=0.0323, \quad \tilde{\chi }_0=0.25, \quad K_\chi=350, \quad K_\rho =96.1.
\end{equation*}
We choose also the initial conditions 
\begin{equation*}
\tau(0)=0,\quad \tilde{\rho}(0)=10^{-6}, \quad \tilde{\chi}(0)=0.18, \quad \beta(0)=0.
\end{equation*}
We take the shear modulus for copper to be $\mu =50000$ MPa. The plots of the resolved shear stress $\tau (\gamma )$ as function of $\gamma $ for copper found by the numerical integration of \eqref{eq:3.5} (bold lines) and \eqref{eq:3.6} (dashed lines) for three different resolved shear rates $0.1/$s (red), $10/$s (blue), and $1000/$s (green), at room temperature $298\,$K, and at $\varphi =30^\circ$ are shown in Fig.~\ref{fig:3}. It can be seen that $\tau(\gamma )$ is rate-sensitive. Besides, the steady-state stresses $\tau_{ss}$ calculated by the finite deformation TDT are somewhat smaller than those computed by the small strain theory. 

\begin{figure}[htb]
	\centering
	\includegraphics[height=5cm]{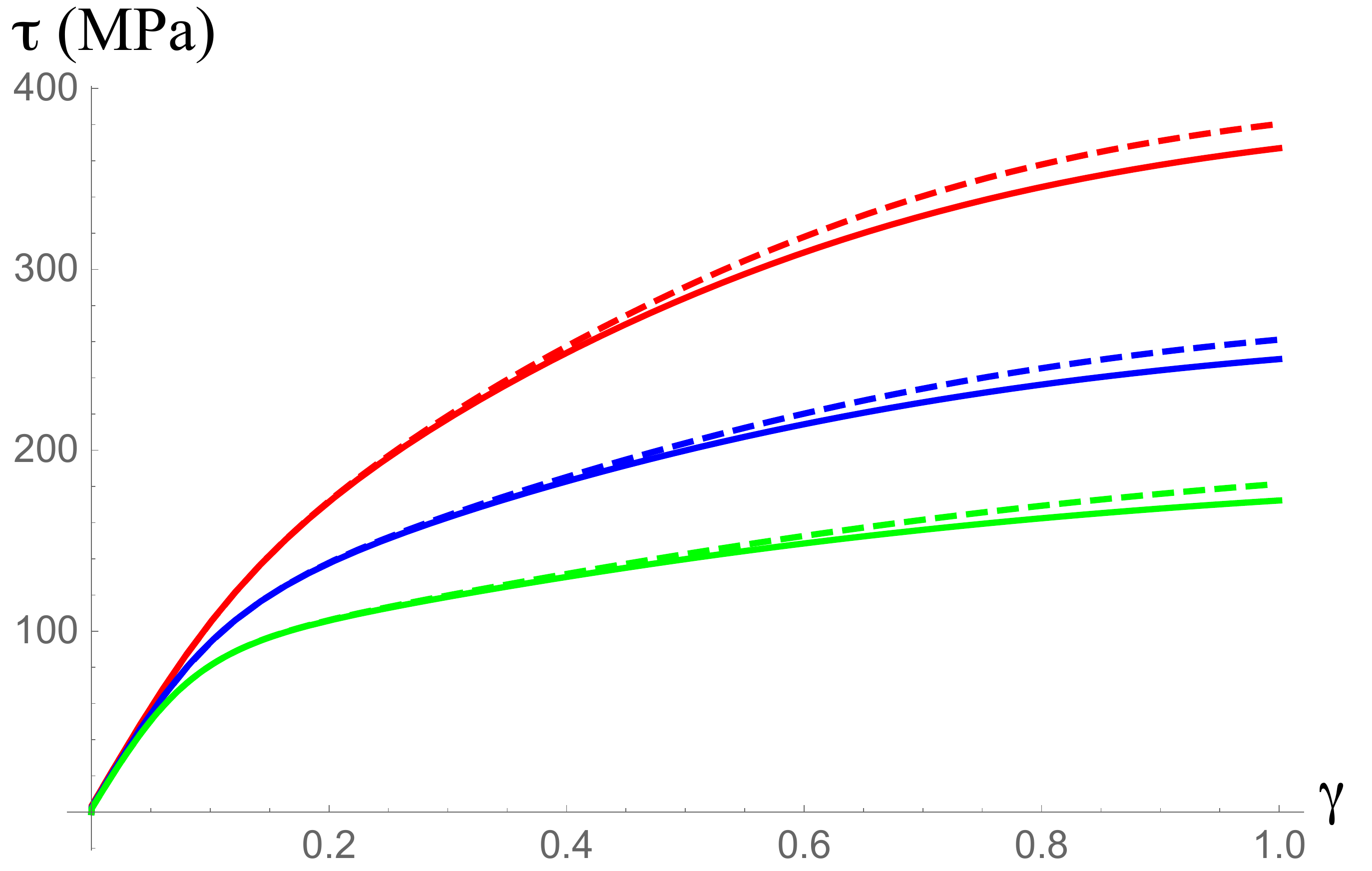}
	\caption{Resolved shear stress $\tau $ (in MPa) versus shear strain $\gamma $ for copper at the shear rate $10/$s, at $\varphi =30^\circ$, and at three temperatures $273\,$K (red), $473\,$K (blue), and $673\,$K (green): (i) finite strain (bold line), (ii) small strain theory (dashed line).}
	\label{fig:4}
\end{figure}

Fig.~\ref{fig:4} plots the resolved shear stress $\tau (\gamma)$ as function of $\gamma $ for copper found by the numerical integration of \eqref{eq:3.5} (bold lines) and \eqref{eq:3.6} (dashed lines) at the shear rate $10/$s, at $\varphi =30^\circ$, and at three temperatures $273\,$K (red), $473\,$K (blue), and $673\,$K (green). We see that the higher the temperature, the lower is the hardening rate and the corresponding steady-state stresses.

\begin{figure}[htb]
	\centering
	\includegraphics[height=5cm]{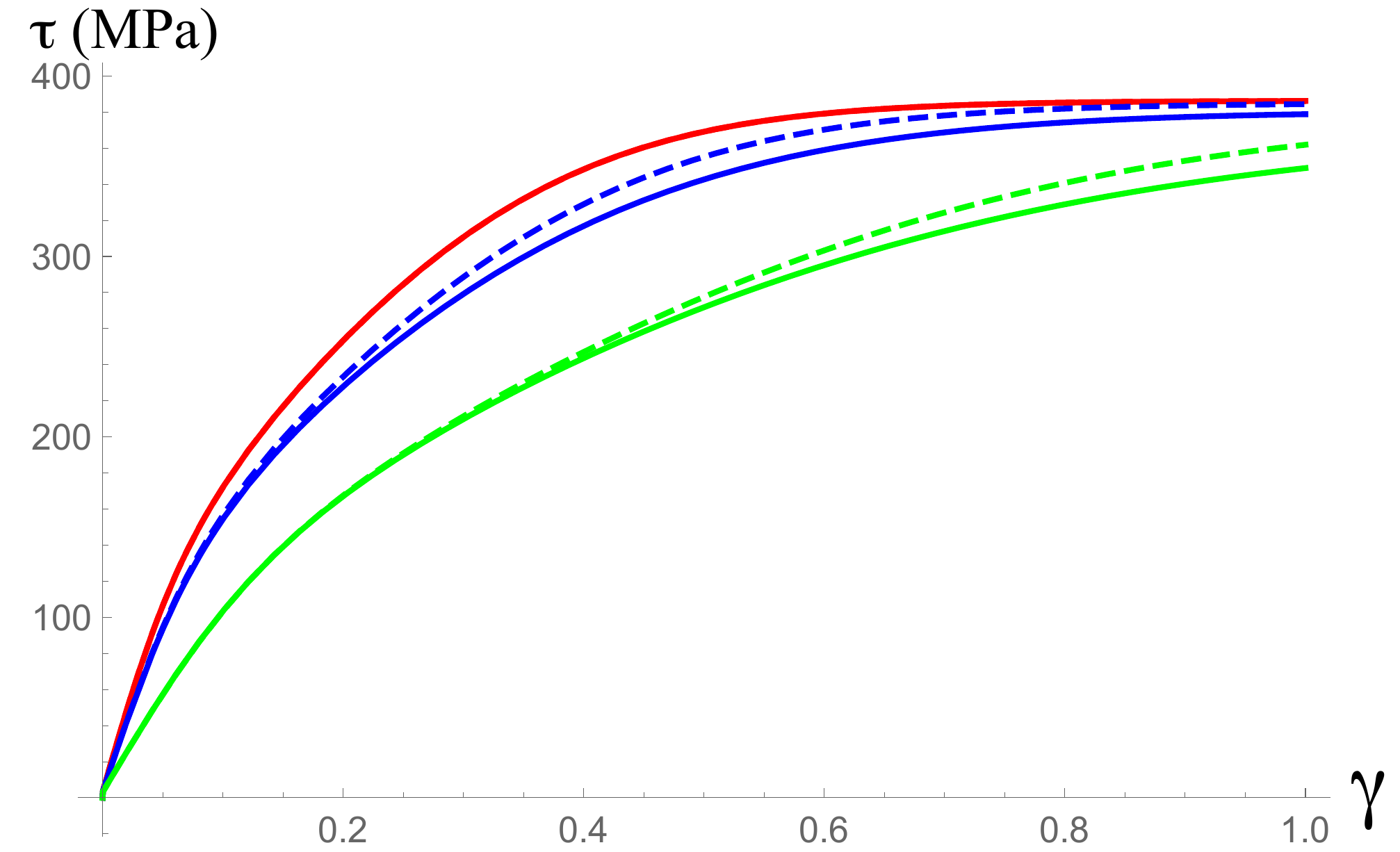}
	\caption{Resolved shear stress $\tau $ (in MPa) versus shear strain $\gamma $ for copper at room temperature $298\,$K, at the shear rate $10/$s, and at three angles of slip $\varphi =0^\circ$ (red), $\varphi =15^\circ$ (blue), and $\varphi =30^\circ$ (green): (i) finite strain (bold line), (ii) small strain theory (dashed line).}
	\label{fig:5}
\end{figure}

Fig.~\ref{fig:5} plots the resolved shear stress $\tau (\gamma)$ as function of $\gamma $ for copper found by the numerical integration of \eqref{eq:3.5} (bold lines) and \eqref{eq:3.6} (dashed lines) at room temperature $298\,$K, at the shear rate $10/$s, and at three angles of slip $\varphi =0^\circ$ (red), $\varphi =15^\circ$ (blue), and $\varphi =30^\circ$ (green). Note that at $\varphi=0^\circ$ the results of small strain and large strain theories coincide. As the angle $\varphi$ increases and approaches $45^\circ$, the resolved shear stress diminishes to zero. The latter becomes negative if the angle $\varphi $ is larger than $45^\circ$. 

\begin{figure}[htb]
	\centering
	\includegraphics[height=5cm]{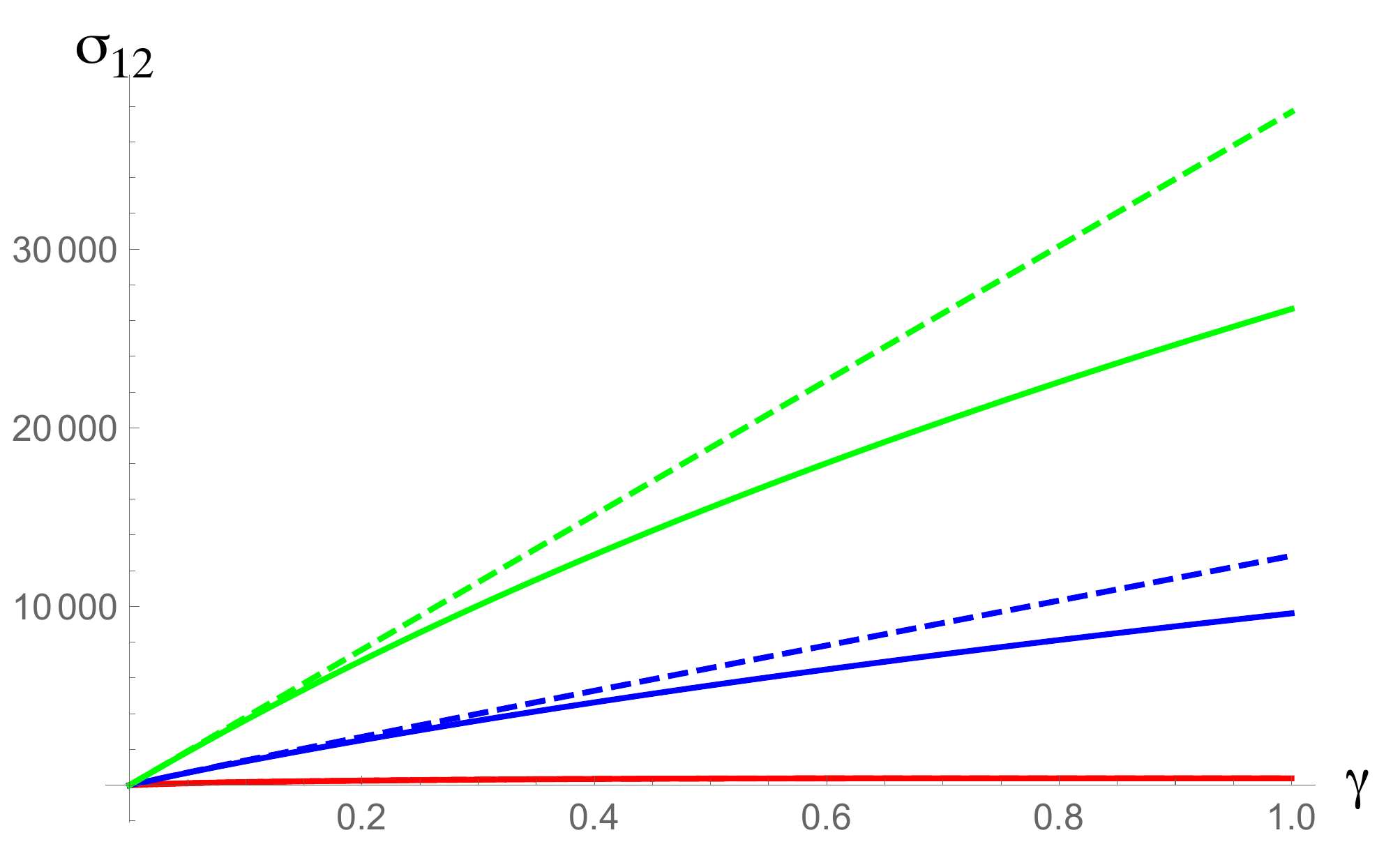}
	\caption{Shear stress $\sigma_{12}$ (in MPa) versus shear strain $\gamma $ for copper at room temperature $298\,$K, at the shear rate $10/$s, and at three angles of slip $\varphi =0^\circ$ (red), $\varphi =15^\circ$ (blue), and $\varphi =30^\circ$ (green): (i) finite strain (bold line), (ii) small strain theory (dashed line).}
	\label{fig:6}
\end{figure}

Although the resolved shear stress decreases with the increasing slip angle (for $\varphi <45^\circ$), the true shear stress $\sigma_{12}$ increases with the increasing slip angle. To see this, let us first compute the true Cauchy stress
\begin{equation*}
\bsigma =J^{-1}\vb{F}^e\cdot \mu (\vb{I}-\vb{C}^{e-1})\cdot \vb{F}^{eT}=\mu (\vb{F}^e\cdot \vb{F}^{eT}-\vb{I}).
\end{equation*}
With \eqref{plastic} we find that
\begin{equation*}
\sigma_{12}=\frac{1}{2}\mu [(2+\beta^2)\gamma -\beta(2+\beta \gamma)\cos 2\varphi +\beta(\beta-2\gamma )\sin 2\varphi ].
\end{equation*}
The small strain theory yields
\begin{equation*}
\sigma_{12}=\mu (\gamma -\beta \cos 2\varphi ).
\end{equation*}
Since
\begin{equation*}
\sigma_{11}=\mu \beta \sin 2\varphi , \quad \sigma_{11}=-\mu \beta \sin 2\varphi ,
\end{equation*}
according to the small strain theory, it is easy to check that the resolved shear stress $\tau =s_i\sigma_{ij}m_j=\mu (\gamma \cos 2\varphi -\beta )$ as expected. Fig.~\ref{fig:6} show the plots of the shear stress $\sigma_{12}(\gamma )$ for copper as function of $\gamma $ computed according to \eqref{eq:3.5} (bold lines) and \eqref{eq:3.6} (dashed lines) at room temperature $298\,$K, at the shear rate $10/$s, and at three angles of slip $\varphi =0^\circ$ (red), $\varphi =15^\circ$ (blue), and $\varphi =30^\circ$ (green). We see clearly that, as $\varphi $ increases the shear stress $\sigma_{12}$ increases too. However, this is valid only for $\varphi < 45^\circ$.

\section{Conclusions and discussions}\label{sec:5}

In this paper, we have developed the extension of the LBL-theory to non-uniform finite plastic deformations for single crystals deforming in single slip. For single crystals deforming in multiple slips, the set of state variables must be enlarged to include multiple densities of dislocation populations, each describing different types of dislocations with different orientations, and we will need separate equations of motion for each of these populations. But even in this case, we can assume that the dislocation populations will retain their identity. Consequently, the generalization to a multi-slip version of the present theory for single crystals should not pose fundamentally new problems. The stress-strain curves of finite strain constrained shear deformation show sensitivity to temperature, strain rate and orientation of the slip system. However, this case study serves only as an illustration of the theory. To compare with real experiments, tension/compression and torsion tests should be considered instead. This and another problem of finite deformation with excess dislocations exhibiting kinematic work hardening and Bauschinger and size effects will be discussed in our forthcoming papers.

\end{document}